# Interactive Sensor Dashboard for Smart Manufacturing

LRD Murthy, Somnath Arjun, Kamalpreet Singh Saluja, Pradipta Biswas
Center of Product Design and Manufacturing
**Indian Institute of Science**
Bangalore, India
lrdmurthy@iisc.ac.in, somnatharjun@iisc.ac.in, kamalpreets@iisc.ac.in, pradipta@iisc.ac.in

*Abstract*—This paper presents development of a smart sensor dashboard for Industry 4.0 encompassing both 2D and 3D visualization modules. In 2D module, we described physical connections among sensors and visualization modules and rendering data on 2D screen. A user study was presented where participants answered a few questions using four types of graphs. We analyzed eye gaze patterns in screen, number of correct answers and response time for all the four graphs. For 3D module, we developed a VR digital twin for sensor data visualization. A user study was presented evaluating the effect of different feedback scenarios on quantitative and qualitative metrics of interaction in the virtual environment. We compared visual and haptic feedback and a multimodal combination of both visual and haptic feedback for VR environment. We found that haptic feedback significantly improved quantitative metrics of interaction than a no feedback case whereas a multimodal feedback is significantly improved qualitative metrics of the interaction.

## I. Introduction

Visualizing and interacting with large scale data is an open problem in the context of information visualization. For example, let us consider a smart factory spread over several buildings and workshops each containing several environment monitoring sensors. Visualizing these sensors' information over a time period and for each sensor separately requires large amount of screen estate. Additionally, users may require interacting with this data to infer about key performance objectives like productivity, pollution level, safety and so on which adds more challenge to the visualization platform. Existing visualization techniques for smart manufacturing explored representing relationship among data through establishing ontologies and visualizing network diagram among different items in [1], [2] and so on. Sackett, Al-Gaylani, Tiwari, & Williams [3] presented a review on existing visualization techniques but did not provide detail on visualizing both temporal and spatial information simultaneously. Existing smart manufacturing set up at Cranfield and Sheffield universities are exploring using state-of-the-art virtual reality, augmented reality and projected displays (e.g. Microsoft HoloLens) mainly for explaining individual components or working principle of complex machines. Existing smart manufacturing processes are investigating large touchscreen (e.g.: Clevertouch Plus) and different virtual reality systems for visualization although those are not exclusively used to visualize sensor data. Existing augmented reality systems mostly use a tablet computer to zoom in small circuit element or showing descriptions of individual components but not for helping sensor fusion or visualization of sensor data.

Our research is addressing large scale sensor visualization and interaction from the following three facets

- Investigating different topology and network options to connect multiple sensors to visualization module.

- Developing 2-dimensional sensor dashboard and evaluating different visualization strategy by analyzing eye gaze of users.

- Developing a 3-D digital twin with embedded IoT modules showing sensor data and investigating multimodal interaction with the virtual reality system.

The paper is organized as follows. The next section presents a brief literature survey followed by description of the three modules presented above. Section IV presents two user studies followed by concluding remarks.

## II. Literature Survey

Sensor data are time-series data that typically have more than three dimensions. Visualizing high-dimensional data is an active research area, and visualization community has been doing research in this field from past three decades. Visualization is part of a large process with sequence of stages that can be studied independently in terms of algorithms, data structures and coordinate systems. The physical limitations of display devices and our visual systems prevent the direct display and rapid recognition of data with dimensions higher than two or three. The visualization process starts with the raw data that will be potentially visualized. In the past, a broad number of approaches like PCA [4], MDS [5] have been introduced to visually convey high-dimensional information by transforming them to low dimensional projections or abstract representations. Next stage of visualization pipeline involves mapping of components of data record to the features of graphical attributes, techniques such as non-projective mapping between n-dimensional and 2-dimensional sets like parallel coordinates[6], use of faces to represent points in k-dimensional space graphically [7] have been introduced in the past. Last stage of visualization is the rendering process that generates images in the screen space, clutter reduction [8] is one of the several efforts that has been made for this stage of visualization. Researches that are focused on high-dimensional data visualization falls in one of the three stages of visualization, for instance, visual data mining [9], quality measures [10,11,12,13,14,15]. Visualization and interactive tools have been developed for identification and understanding of clusters and complex patterns in high-dimensional data. SMARTexplore [16] is one such example which introduces a novel visual analytics technique that simplifies the identification and understanding of clusters, correlations and complex patterns in high-dimensional data.

In recent time with advancement of interactive devices, the visualization process also changed from a one-way rendering system to an interactive system allowing users to interact with data to interpret it at various stages of aggregation. For example, in our daily life, we interact with a map application in our smartphone through multiple modalities like multi-touch, gesture, speech recognition and so on. Richard Bolt's "Put That There" [17] system first demonstrated the importance of multimodal interfaces. Later, numerous systems [18,19,20] followed which integrated various input modalities like speech, gesture, pen input etc., Even though the modalities like speech, graphics, virtual character are explored for output, relatively less work has happened on the multimodal output front, which focuses on how the systems should respond to user actions. In case of Virtual reality, where an entire virtual environment can provide immersive experience visually, other sensations that humans can feel in real world like tactile responses, environmental sensations like temperature and pressure are inherently not present. Hence, the design of multimodal output in virtual reality must be studied with wide range of contexts and applications. Erler [21] studied effect of haptic and visual feedback on a task involving grabbing and throwing a VR basketball game using the controller. It was found that presence of feedback compared to no-feedback improved accuracy and predictability, but no difference was reported in task load and usability. They also noted that releasing a button on the controller is a bad metaphor for releasing a grabbed virtual object.

### III. PROPOSED APPROACH

#### A. IOT Nodes

Our proposed visualization system (Figure 1) consists of the following three parts
- IoT unit with a single-board computer
- An interactive visualization module
- An early warning system

Each IoT node consist of a single-board computer and a set of sensors. The single board computer records data from different sensors and fuse sensor signals, when required. Presently, we developed the IoT node for environment tracking using MQ-5 Smoke Sensor, HL-83 Flood / Rain Sensor and a DS18B20 temperature sensor. The visualization module runs on a standard desktop or laptop computer attached to a screen. The next section describes the different 2D visualization techniques explored so far. The visualization module is integrated to an analytics and alert module that constantly analyses data recorded from sensors and set out an alert if any values cross a threshold. The user can send either a manual alert or set up an automatic alert for one, all or a subset of sensors using the graphical user interface provided with the visualization module. The alert is sent as a Gmail to a pre-recorded email address. We chose Gmail as alert platform as sending a SMS requires the visualization module to be integrated with a telephone communication unit, while social networking messages (like Facebook or WhatsApp alert) require the end user to subscribe to a social networking site. The Gmail can be received on a smartphone or smartwatch. The Gmail message summarizes sensor readings and has a subject stating the type of sensor creating the alert. More details on the system can be found in a separate paper [22] and a video demonstration of the system can be found at https://youtu.be/leRrcdKsyPM .

#### B. 2D Visualization

Initially, we undertook a study to investigate how users interpret information from different types of graphs [23]. A software consisting of four visualization techniques and a set of five questions with multiple choice answers were developed. Users were asked to answer the set of questions as a part of the task. Gaze locations was captured while performing the task to analyze user's attention on parts of the graphs using eye tracker. Analysis of gaze data has been done based on soft clustering using Expectation Maximization algorithm. Results of the study shows that bar graph has the highest accuracy and area graph has the lowest response time while performing task. Based on the results, we developed a two-dimensional interface that can help to visualize high dimensional data with high accuracy and lower response time. In our new visualization system, circle of the scatter plots represents the area where the sensors are located, and the bars represent the values of the sensors over a period of time. Figure 2 shows a sample interface of the software developed for undertaking the study and new 2D visualization technique. Presently, we are collecting data on a user study to compare this visualization with a 3D visualization described in next section.

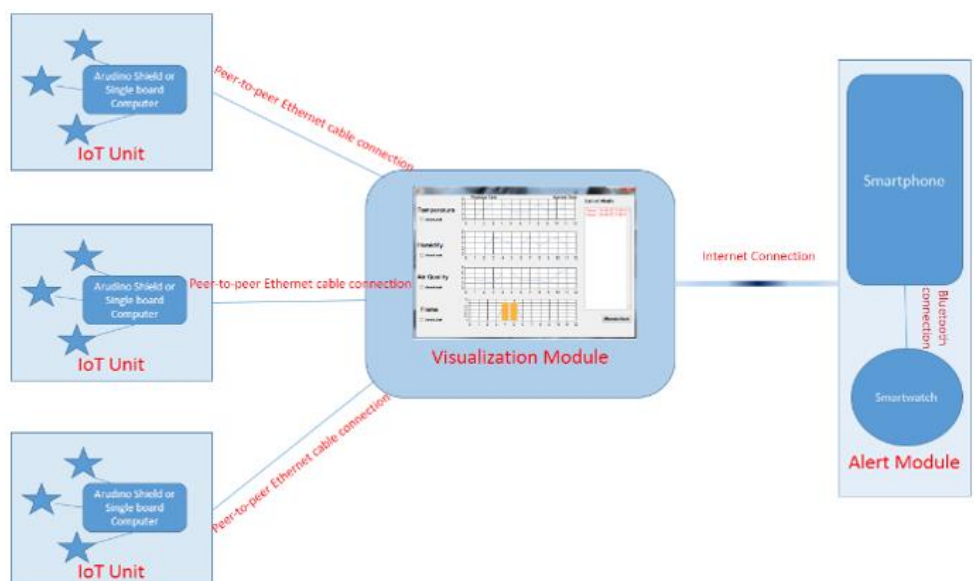

Fig. 1 Different Modules of the System

### C. VR Digital twin

We assumed that the 3D visualization will provide an extra dimension to the view and allows users interact in a natural way. We aimed to study and understand its efficacy when compared to 2D visualization. We chose virtual reality as the platform for creating a 3D digital twin for a smart factory. A walk-through the large manufacturing setup, across all floors would also try to address the interaction and the screen estate issues mentioned earlier in addition to providing 3D visualization.

Before we can investigate the visualization paradigm in 2D and 3D, there is a need for understanding the interaction paradigm in 3D. Even though VR provides immersive experience, unlike real world it does not provide any physical feedback to the user. So, the design of the responses from the system for every interaction, in other words, the effect of modalities of output of the VR interactions on the user performance must be studied. For this, as a part of our study, we chose *"No feedback case"* as the baseline and this has been compared with visual, haptic and multimodal output involving both visual and haptic feedbacks. We developed a 3D version of the Fitts' Law [24] task as it has been traditionally used to compare different modalities of interaction in direct manipulation interfaces. Twelve participants undertook the trial with seven indices of difficulty in each feedback case. The details of the user study are explained in the Section IV *B*. Results shows that users prefer the multimodal output. Next, we have developed a VR model of our own lab and set up visualization graphics at the locations of IoT nodes to embed real-time sensor readings on the virtual layout. Users can browse through the virtual set up using 3D glass and as they touch any of the visualization, it provides both visual and haptic feedback based on sensor readings.

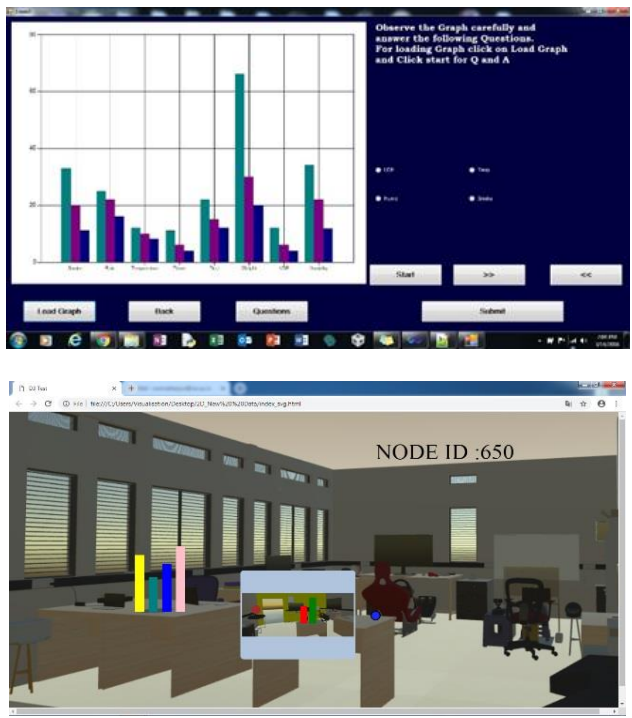

Fig. 2 Bar graph and two-dimensional visualisation

We integrated ambient light sensor (BH1750) and, temperature and humidity sensor (DHT22) to show real-time visualization of data stream(s) in VR setup. Both sensors

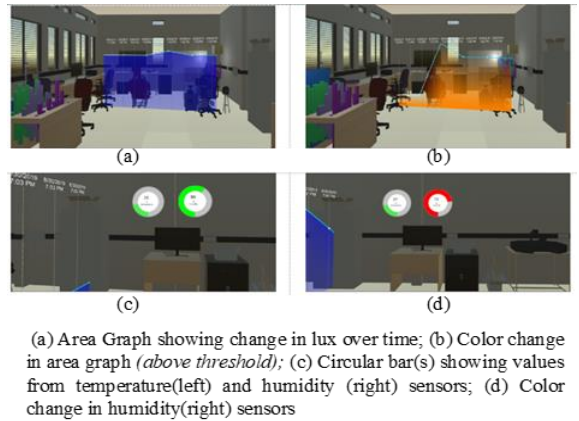

(a) Area Graph showing change in lux over time; (b) Color change in area graph *(above threshold)*; (c) Circular bar(s) showing values from temperature(left) and humidity (right) sensors; (d) Color change in humidity(right) sensors provide digital output. The BH1750 Sensor has a built-in 16-bit A2D converter and output unit is lux. The DHT22 sensor provides temperature in celcius and humidity as relative percentage.

Fig. 3 Visualization of real-time sensor data inside VR Digital Twin

Sensors are interfaced to the VR machine through their respective wireless module(s) [25]. After establishing a peer-to-peer connection, individual wireless module communicates with VR machine using UDP protocol at a frequency of 1 Hz.

The data stream taken from light sensor is visualized as an area graph [26] (Figure 3a). The graph shows change in light intensity over time. Data obtained from temperature and humidity sensor is shown as separate circular bar (Figure 3c). The color of the area graph and the circular bar(s) is changed if the value exceeds a threshold (Figure(s) 3b and 3d). We also provided haptic feedback on the hands for the same.

A video demonstration of the system can be found at https://youtu.be/-UOB9Stgxd8 .

## IV. USER STUDIES

### A. Study On Graph Visualization

**Aim of the Study:** We undertook the following study to investigate how users interpret information from different types of graphs. Similar kind of study was earlier conducted [27], where gaze locations was captured to analyse user's attention to specific parts of the visualization technique. We have developed new techniques for analysing gaze data based on soft clustering. In particular, we investigated Expectation Maximization (EM) algorithm. We have used XB cluster validation index [10] for validating optimum number of clusters. Using these soft clustering techniques, we could automatically identify the number and locations of areas of interest in a visual display. This study will be useful to point the anomalies in the current visualization techniques.

**Expectation–Maximization** (EM) is an iterative method to find Maximum Likelihood or Maximum a Posteriori (MAP) estimates of parameters in statistical models, where the model depends on unobserved latent variables. The EM iteration alternates between performing an expectation (E) step, which creates a function for the expectation of the log-likelihood evaluated using the current estimate for the parameters, and a maximization (M) step, which computes parameters maximizing the expected log-likelihood found on the *E* step. These parameter-estimates are then used to determine the distribution of the latent variables in the next E step.

**XB Cluster Validation Index:** A cluster validity function proposed by Xie and Beni [10] is used to evaluate the fitness of partitions produced by clustering algorithms. It is defined as the ratio of the compactness measure and separation measure, i.e. lower index value indicates fitter partitions.

In the following paragraphs, we described detail of the study.

**Participants:** We conducted the user study with 9 participants, among them 6 males and 3 females, everyone between 20 and 35 years.

**Materials:** We used a Tobii Eye-X tracker for recording eye gaze, 29 inch display monitor with 1366×768 screen resolution and Lenovo yoga laptop with i5 processor for conducting the user study. As per Tobii Eye-X license agreement, we did not store raw eye gaze data, rather only analyzed data in real time.

**Design:** We developed a software which consists of four visualization techniques and five set of questions with multiple choice answers. An eye gaze tracker was placed at the bottom of the screen. Participants were asked to seat at 75 cm away from the screen. They were instructed to answer a set of questions by investigating the graph. Figure 4a and 4b below shows a sample interface of the system. We used the following four types of graphs

- Bar Graph
- Line Graph
- Radar Graph
- Area Graph

For each graph, the following set of questions were displayed one at a time. The order of presentations of the graphs and questions was randomized to reduce learning and order effects.

**Q1:** How many sensors have lesser average value than average of all low values?

**Q2:** Average of which sensor is approximately same as the average of all sensor's average value?

**Q3:** Sensor having high value greater than 50 and less than 100?

**Q4:** Two sensor reading showing nearly equal low values with minimum difference?

**Q5:** What is the approximate average of all High values of sensors?

**Procedure:** Participants were briefed about the experiment. For each participant, the eye tracker was calibrated using the Tobii calibration routine [28]. After calibration, participants were asked to undertake the study. The X-Y coordinates of the gaze location and response to each question with timestamp were logged in a text file.

**Results and Discussion:** Our analysis found significant difference in gaze and response behaviour for different graphs while performing similar tasks. We analysed four dependent variables - number of correct answers, average time for correct answers, total time taken for all answers and the optimal number of clusters for user's gaze fixation. Moreover, we analysed user's gaze fixation using Expectation Maximization algorithm, which indicates that more fixations mean more eye gaze movement requiring longer duration to analyse data. We found that Bar graph has the highest number of correct answers with 27 correct answers and Radar graph has the lowest with 19 correct answers out of 45 questions, all users cumulatively. Area Graph has the lowest average response time for individual questions (24.6 seconds) and total time for all questions (155.86 seconds). We also noticed that the average response time of Bar graph is high and number of correct answers for Area graph is low. This speed-accuracy trade-off may lead us to future research questions. We undertook one-way ANOVA (Analysis of Variance) for all dependent variables and did not find significant differences for any of the dependent variables, [p>0.05].

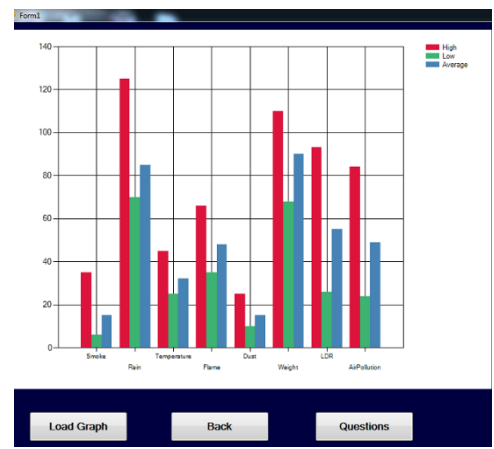

Fig. 4a: Graphical portion of the System

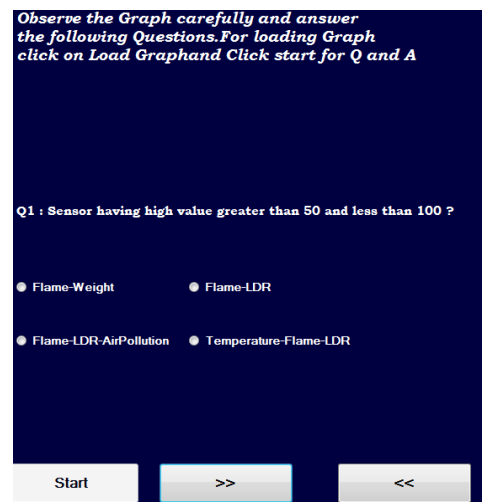

Fig. 4b: Question and Answer portion of the System

Table I: Values of dependent variables

|  | Bar | Line | Radar | Area |
|---|---|---|---|---|
| CA | 3 | 2.22 | 2.11 | 2.67 |
| ART (secs) | 37.33 | 30.35 | 39.45 | 24.6 |
| TRT (secs) | 196.77 | 166.35 | 212.83 | 155.86 |
| ONC | 4.55 | 4.66 | 3.55 | 3.55 |

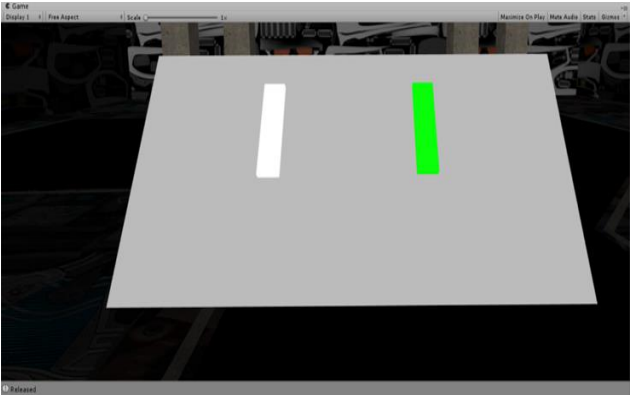

Fig. 5 Fitts' Law Study in VR Setup for Visual Feedback Case

CA: Number of correct answers
ART: Average Response time for correct answers in seconds
TRT: Total response time for all answers in seconds
ONC: Optimum number of clusters

Table 1 above shows average values of all dependent variables for all types of graphs. We analysed the sequences of eye gaze fixations in all four graphs like Steichen's study [27]. Our initial analysis consists of dividing the screen in 9 regions and analysing first saccade positions and subsequent gaze movements from initial position. We noted, user's eye gaze first fixated on the central part and moving to bottom-centre of the graph subsequently for all types of graph. We also noticed that initial patterns for eye gaze fixations are similar for every graph however subsequent gaze movements are different for each graph. We found after mining those movements that the sequence of bottom-centre to middle-centre and top-right to middle-centre is most frequent among all two-region sequences.

*B. Study on Effect of Feedback in VR*

The previous case study involves optimizing visual rendering of the 2D sensor dashboard while the second case study compared visual, haptic and multimodal feedback in a VR environment. We designed a study inspired by the experiment reported by Fitts [24].

**Participants:** 12 participants (9 male and 3 females; Age range: 23 – 33 years; All right-handed) are recruited for the study. Each prospective participant answered the simulator sickness questionnaire [29]. Even though two participants wanted slightly longer breaks between the study compared to the others, simulator sickness is not a problem in our study. A goniometer is used to verify the degree of hand movement and all the participants have full range movement.

**Materials:** We used Unity [30] to design the virtual environment and Oculus Rift [31] system for displaying it. The tracking sensors of Oculus Rift are kept undisturbed throughout the study. We used a desktop with i7 processor and Nvidia GTX 1080 Ti for this study. Oculus rift right hand controller is used for performing the task.

**Design:** We created a virtual environment that resembles the original Fitts' experiment [24]. The environment contains a platform tilted towards the participant. The platform contains two cubes, equidistant from the centre. A blue circle has shown as a representation of the rift controller in the 3D space. We considered 3 different values for distance between the two cubes (A = 1, 1.5 and 2) and 3 different values for the width of the cubes (0.05, 0.1, 0.15) on platform and hence we have 9 study cases. The index of difficulty (ID) is calculated based on the following formula.

$$ID = log_2\left(\frac{A}{W} + 1\right)$$

The numbers above mentioned have the units same as the default Unity settings. Based on the above choice of distance and widths, we obtained 7 Index of difficulties. Using ID and measured mean movement time for each case, we compute throughput as the ratio of Index of difficulty and mean movement time.

The experiment is conducted with 4 feedback cases.
1. No feedback to the participant
2. Visual Feedback: The cube turns green as the participant reaches it.
3. Haptic feedback is provided as the participant reaches a cube.
   a. Vibrational Haptic feedback is given using the oculus controller and the in-built classes OVRHaptics and OVRHapticsClip (*Haptics*) [32].
4. Both visual and haptic feedback is given once the participant reaches the cube.

The participants were instructed to
- Bring the controller onto either of the cube on the platform.
- Press "A" button on the controller, drag the controller towards the other cube.
- Release the button.

Movement time in each case is the time between a 'press' and 'release' of the button 'A' on the controller. For a given width of the cube, distance between the cubes and a feedback case, a participant does the above-mentioned task for 25 times. Mean movement time is considered as the average of the movement times over those 25 measurements. If the participant presses or releases the button outside the boundary of the cubes on the platform, it is considered as an error and is measured from the nearest edge of the cube.

**Procedure:** Each participant is instructed to get accustomed with the VR environment, as 8 out of 12 participants experiencing VR for the first time. Then, they answered simulator sickness questionnaire to make sure that they do not feel uncomfortable during the experiment. The participants are asked to perform the above-mentioned task in VR version of Fitts' experimental set up for around 15 minutes. This is to minimize the learning of the task during the actual trial. Since we considered 3 widths for the cube, 3 distances between the cubes and 4 feedback cases, we get a total of 36 test cases. These 36 test cases are randomized for each participant. All participants are instructed to do the task "*as quickly and as accurately as possible*". In each trial, mean movement time and errors committed are recorded. Figure 5 shows the VR environment for visual feedback case. After each participant completes their trial, subjective feedback is collected using NASA TLX for cognitive load and SUS questionnaire for subjective preference.

Table II: Quantitative and Qualitative Results of VR Fitts' Study

| **Feedback Type** | **Mean Movement Time (ms)** | **Mean Throughput (bits/sec)** | **Mean Error** | **Mean SUS** | **Mean TLX** |
|---|---|---|---|---|---|
| No Feedback | 665.01 (94.52) | 6.107 (0.48) | 1.07 (0.28) | 58.13 (12.9) | 49.41 (18.1) |
| Visual | 670.39 (96.74) | 6.05 (0.53) | 1.09 (0.28) | 75.42 (14.7) | 36.94 (18.1) |
| Haptic | 613.23 (86.67) | 6.64 (0.44) | 1.06 (0.29) | 65.41 (14.6) | 43.26 (16.2) |
| Multimodal (Visual and Haptic) | 641.65 (87.02) | 6.33 (0.52) | 1.06 (0.28) | 78.33 (13) | 30.11 (16.6) |

**Results and Discussion:** Initially, we calculated the average movement time of hand for each modality. Figure 6 plots the mean movement time with respect to indices of difficulty in each feedback case. In addition to that, the figure also plots the least square fit (LS) line for each feedback case. Except visual feedback case, we can observe that the R square values (coefficient of determination) [33] is around 0.9 in all other feedback cases. Lowest movement time was found for Haptic feedback.

We undertook one-way ANOVA on movement times but did not find any significant effect. Six pairwise t-tests found that

- Visual and No-feedback cases needed significantly higher (p<0.05) movement times than haptic and multimodal feedback cases.
- Haptic feedback needed significantly lower time than multimodal feedback case.
- There was no significant difference between visual and no-feedback cases.

We undertook one-way ANOVA on TLX scores but did not find any significant effect. Six pairwise t-tests found that

- In Visual and Multimodal cases, participants have experienced significantly lower (p<0.05) cognitive load than no feedback case.
- Participants experienced significantly lower cognitive load in multimodal feedback case than in visual and haptic feedback cases.
- The experienced cognitive load in haptic feedback case is not significantly different from no-feedback and visual feedback cases.

An one way ANOVA for SUS scores found significant difference among different feedback cases F(3,44)=5.4, p<0.05, η2 = 0.27. Six pairwise t-tests found that

- Subjective preference to Visual and Multimodal feedback cases is significantly higher (p<0.05) than to a No Feedback case.
- Preference towards multimodal feedback is significantly higher than to haptic feedback.
- Preference towards a multimodal feedback than to a visual feedback is not significantly different.
- Preference towards no feedback and visual feedback is not significantly different from a haptic feedback case.

The grey shaded regions in the Table II signify the cases with statistically significant improvement compared to the baseline no-feedback case.

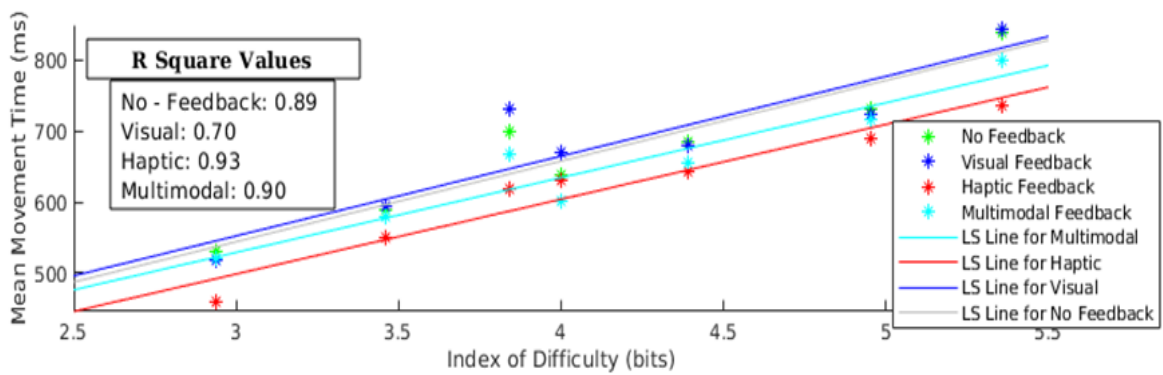

Fig. 6 Mean Movement Time Vs Index of Difficulty for various feedback cases augmented with a least-squares line

From the results, we infer that quantitatively haptic and multimodal have higher throughput and qualitatively multimodal and visual feedback is preferred by users.

## V. Conclusions and Future Work

This paper presents development of a smart sensor dashboard having both 2D and 3D visualization modules. It describes different modules of the sensor dashboard and presents two user studies on analyzing existing visualization techniques in 2D. We also studied various feedback cases in the case of 3D virtual sensor dashboard.

From the gaze fixation analysis for existing visualization techniques in 2D, we observed that initial patterns for eye gaze fixations are similar for every graph however subsequent gaze movements are different for each graph. We found after mining those movements that the sequence of bottom-centre to middle-centre and top-right to middle-centre is most frequent among all two-region sequences.

In the case of VR sensor dashboard study, we studied interaction paradigm across 4 feedback cases. We observed that the users perceived multimodal feedback consisting of visual and haptic stimuli better on both qualitative and quantitative analysis. We take these results forward for interaction design in the case of 3D visualization. We shall study the visualization in 3D, and we will compare its efficacy against the results we obtained for 2D visualization scenario. Since the walk-through of a virtual room with live sensor nodes is already designed, we shall carry out experiments on evaluating the efficacy of these dynamic sensor dash boards against static screen dashboards. The application we envisage as a result of our future work at this moment are a virtual smart factory application using which a person can inspect and interact with every entity of the virtual smart factory to retrieve information and to discharge commands.